\documentstyle[psfig,12pt]{article}
\begin{document}
\baselineskip 0.0 cm
\title{\sc The effect of non-radial motions on the CDM model predictions}
\author{{\bf A. Del Popolo} ~ and ~ {\bf M. Gambera}}
\date{}
\maketitle
\begin{center}
{\sc Istituto di Astronomia dell'Universit\`a di Catania,\\ 
Viale A.Doria, 6 - I 95125 Catania, ITALY \\}
~\\

{\bf To be published:}\\
In the Proceedings of {\it " The VIII Conference on Theoretical Physics:\\ 
General Relativity and Gravitation "} \\
BISTRITZA - JUNE 15-18, 1998 - Romania\\
\end{center}
~\\
\begin{abstract}
In this paper we show how non-radial motions, originating 
from the tidal interaction of the irregular mass distribution 
within and around protoclusters, can solve some of the problems of 
the CDM model. Firstly the discrepancy between the CDM predicted 
two-points correlation function of clusters and the observed one.  
We compare the two-points correlation function, that we obtain taking 
account of non-radial motions, with that obtained by Sutherland \& Efstathiou
(1991) from the analysis of Geller \& Hucra's (1988) deep redshift survey 
and with the data points for the APM clusters obtained by 
Efstathiou et al. (1992). Secondly the problem of the X-ray clusters abundance 
over-production predicted by the CDM model. In this case we compare   
the X-ray temperature distribution function, calculated using
Press-Schechter theory and Evrard's (1990) prescriptions for the 
mass-temperature
relation, taking also account of the non-radial motions, with Henry \& 
Arnaud (1991) and Edge et al. (1990)  
X-ray temperature distributions for local clusters. 
We find that in both cases the model is in good agreement with experimental data.
Finally we calculate the bias coefficient using a selection function
that takes into account the effects of non-radial motions, and we show that 
the {\it bias} so obtained can account for a substantial part of the total 
bias required by observations on cluster scales.\\
~\\
keywords: ~ {\it cosmology: theory-large scale structure of Universe - galaxies: formation}

\end{abstract}
\newpage

\section{Introduction}

Although at his appearence the standard form
of CDM was very successful in describing the observed structures in the
universe (galaxy clustering statistics, structure formation epochs, peculiar
velocity flows) (Peebles, 1982; Blumenthal et al. 1984; Bardeen et al. 1986;
White et al. 1987; Frenk et al. 1988; Efstathiou 1990) recent measurements
have shown several deficiencies of the model, at least if any bias of the
distribution of galaxies relative to the mass is constant with scale. Some
of the most difficult problems to reconcile with the theory  
are 
the
strong
clustering of rich clusters of galaxies, $\xi _{cc}(r)\simeq $ $\left( \frac
r{25h^{-1}Mpc}\right) ^{-2}$ , far in excess of CDM\ predictions (Bahcall \&
Soneira 1983), the X-ray temperature distribution function of clusters,
over-producing the observed cluster abundances (Bartlett \& Silk 1993), the
conflict between the normalisation of the spectrum of the perturbation which
is required by different types of observations. \\ 
Alternative models with more large-scale power than CDM have been introduced
in order to solve the latter problem. Several authors (Peebles 1984;
Efstathiou, Sutherland \& Maddox 1990, Turner 1991,
White, Efstathiou \& Frenk 1993) have lowered the matter density under
the critical value 
($\Omega _m<1$) and they have added a cosmological constant in order to retain
a flat universe ($\Omega _m+\Omega _\Lambda =1$) .
The spectrum of the matter density is specified by the
transfer function, but its shape is
affected because of the fact that the epoch of
matter-radiation equality (characterized by a red-shift $z_{eq}$) is earlier,
$1+z_{eq}$ being increased by a factor $1/\Omega_{m}$. 
Around the epoch $z_\Lambda $, where 
$z_\Lambda=(\Omega_m/\Omega_\Lambda)^{1/3}-1$, 
the growth of the
density contrast slows down and ceases after $z_\Lambda $.
As a consequence the
normalisation of the transfer function begins to fall, even if its shape
is retained.
Mixed dark matter models (MDM) (Bond et al. 1980; Shafi \& Stecker
1984; Valdarnini \& Bonometto 1985; Schaefer et al. 1989; Holtzman 1989;
Schaefer 1991; Shaefer and Shafi 1993; Holtzman \& Primack 1993) increase
the large-scale power because neutrinos free-streaming damps the power on
small scales. Alternatively changing the primeval spectrum
several problems of CDM are
solved (Cen et al. 1992). Finally it is possible to assume
that the threshold for galaxy formation
is not spatially invariant but weakly modulated ($2\%-3\%$ on scales $%
r>10h^{-1}Mpc$) by large scale density fluctuations, with 
the result that the clustering on
large-scale is significantly increased (Bower et al. 1993). \\
Here we propose a different solution to several of the CDM model 
problems connected to the non-radial motions developing during
the protocluster evolution.\\
It has long been
speculated that angular momentum could have a
fundamental role in determining the fate of collapsing proto-structures
and several models have been proposed in which the galaxy
type can be correlated with the angular momentum per unit mass of the
structure itself (Faber 1982; Kashlinsky 1982; Fall 1983).
Some authors (see
Barrow \& Silk 1981, Szalay \& Silk 1983 and Peebles 1990) have proposed that
non-radial motions would be expected within a developing proto-cluster due to
the tidal interaction of the irregular mass distribution around
them, typical of hierarchical clustering models, with the
neighboring proto-clusters.
The kinetic energy of this non-radial motions opposes the collapse
of the proto-cluster, enabling the same to reach statistical equilibrium
before the final collapse (the so called previrialization conjecture by
Davis \& Peebles 1977, Peebles 1990). This effect may prevent the
increase of the slope of the mass autocorrelation function at
separations given by $\xi(r,t) \simeq 1$, expected in the scaling solution for
the growth of $\xi(r,t)$ but not observed in the galaxy two-point
correlation function. The role of non-radial motions has been 
pointed by several authors (see Davis \& Peebles 1983, Gorski 1988, 
Groth et al. 1989, Mo et al. 1993, Weygaert \& Babul 1994, Marzke 
et al. 1995 and Antonuccio \& Colafrancesco 1997). 
Antonuccio \& Colafrancesco 
derived the conditional probability distribution $f_{pk}({\bf v}| \nu)$  
of the peculiar velocity around a peak of a Gaussian density field and 
used the moments of the velocity distribution to study the 
velocity dispersion around the peak.
They showed that regions of the proto-clusters at radii, $ r$, 
greater than the filtering length, $ R_{f}$, contain predominantly 
non-radial motions. \\
Non-radial motions change the energetics of the collapse model
by introducing another
potential energy term. In other words one expects that non-radial motions
change the characteristics of the collapse and in particular the
{\it turn around} epoch, $t_{m}$,
and consequently the critical threshold, $ \delta_{c}$, for collapse.
One expects that non-radial motions produce firstly
a change in the turn around epoch, secondly a new functional
form for $ \delta_{c}$, thirdly
a change of the mass function calculable with the Press-Schechter
(1974) formula and consequently of the predicted X-ray temperature 
distribution function of clusters and finally a modification of the two-point correlation function.
Moreover this study of the role of non-radial motions in
the collapse of density perturbations can help us to give a deeper insight
on the so called problem of biasing. As pointed out by Davis et al. (1985),
unbiased CDM suffers of several problems: pairwise velocity dispersion
larger than the observed one, galaxy correlation function steeper than
observed (see Liddle \& Lyth 1993 and Strauss \& Willick 1995). 
The remedy to these problems is the concept of biasing (Kaiser 1984), 
i.e. that galaxies are more strongly clustered than
the mass distribution from which they originated.
The physical origin of such biasing is not yet clear even if several
mechanisms have been proposed (Rees 1985; Dekel \& Rees 1987; Dekel \& Silk
1986; Carlberg 1991; Cen \& Ostriker 1992; Bower et al. 1993; Silk \& Wyse
1993). Recently Colafrancesco, Antonuccio \& Del Popolo (1995, hereafter CAD) 
have shown
that dynamical friction delays the collapse of low-$\nu$ peaks inducing a 
bias of dynamical nature. Because of dynamical friction under-dense regions
in clusters (the clusters outskirts) accrete less mass with respect 
to that accreted in absence of this dissipative effect and as a consequence 
over-dense regions are biased toward higher mass (Antonuccio \& Colafrancesco 
1995 and Del Popolo \& Gambera, 1996).
Non-radial motions acts in a 
similar fashion to dynamical friction: they delay the shell collapse
consequently inducing a dynamical bias similar to that produced by dynamical
friction. This dynamical bias can be evaluated defining a selection function
similar to that given in CAD and
using Bardeen, Bond, Szalay and Kaiser (1986, hereafter BBKS) prescriptions.
The plan of the paper is the following: in \S 2 we obtain the total
specific angular momentum acquired during expansion by a proto-cluster.
In \S 3 we find the effect of non-radial motion on the 
critical density threshold, $\delta_c$.  
In \S 4 we derive a selection function for the peaks
giving rise to proto-structures while in \S 5 we calculate some values for 
the bias parameter, using the selection function derived, on three relevant 
filtering scales. 
In \S 6 we find the effects 
of non-radial motions 
on the X-ray temperature distribution function and then we compare
this prevision to the X-ray observed data. In \S ~7 we study how 
non-radial motions influence the cluters two-points correlation function. 
\S 8 is devoted to conclusions and discussions. \\

\section{Tidal torques in clusters evolution.}

The explanation of galaxies spins gain through tidal torques was pioneered by
Hoyle (1949) in the context of a collapsing protogalaxy. Peebles (1969)
considered the process in the context of an expanding world model showing
that the angular momentum gained by the matter in a random comoving {\it %
Eulerian} sphere grows at second order in proportion to $t^{5/3}$ (in a
Einstein-de Sitter universe), since when the proto-galaxy is still a small
perturbation, while in the nonlinear stage the growth rate of an oblate
homogeneous spheroid decreases with time as $t^{-1}$. More recent analytic
computations (White 1984, Hoffman 1986, Ryden 1988) and numerical
simulations (Barnes \& Efstathiou 1987) have re-investigated the role of
tidal torques in originating galaxies angular momentum. In particular White
(1984) expanded an analysis by Doroshkevich (1970) showing that the angular
momentum of galaxies grows to first order in proportion to $t$ and that
the  result of Peebles is a consequence of the spherical symmetry imposed in 
the model. White showed that the angular momentum of a Lagrangian sphere does
not grow either in first or in second order while the angular momentum of a
non-spherical volume grows to first order in agreement to Doroshkevich's
result. Hoffman (1986) has been much more involved in the analysis of the
correlation of the growth of angular momentum with the density perturbation $
\delta(r) $. He found an angular momentum-density anticorrelation: high density
peaks acquire less angular momentum than low density peaks. One way to study
the variation of angular momentum with radius in a galaxy is that followed
by Ryden (1988). In this approach the protogalaxy is divided into a series
of mass shells and the torque on each mass shell is computed separately. The
density profile of each proto-structure is approximated by the superposition
of a spherical profile, $\delta (r)$, and a random CDM distribution, ${\bf %
\varepsilon (r)}$, which provides the quadrupole moment of the protogalaxy.
To first order, the initial density can be represented by:
\begin{equation}
\rho ({\bf r})=\rho _b\left[ 1+\delta (r)\right] \left[ 1+\varepsilon ({\bf 
r})\right] 
\end{equation}
where $\rho_{b}$ is the background density and $ \varepsilon(\bf r)$
is given by:
\begin{equation}
\langle |\varepsilon _k|^2 \rangle = P(k)
\end{equation}
being $ P(k)$ the power spectrum, while the density profile 
is (Ryden \& Gunn 1987):
\begin{equation}
\langle \delta (r) \rangle =\frac{\nu \xi (r)}{\xi (0)^{1/2}}-\frac{\vartheta (\nu
\gamma ,\gamma )}{\gamma (1-\gamma ^2)}\left[ \gamma ^2\xi (r)+\frac{%
R_{\ast }^2}3\nabla ^2\xi \right] \cdot \xi (0)^{-1/2} 
\label{eq:dens}
\end{equation}
where $\nu $ is the height of a density peak, $\xi (r)$ is the two points
correlation function, $\gamma $ and $R_{\ast}$  are two spectral parameters 
(BBKS, Eq. 4.6a, 4.6d) while $ \vartheta (\gamma \nu ,\gamma )$ is a function 
given in BBKS (Eq. 6.14).
As shown by Ryden (1988) the net rms torque on a mass shell centered on the
origin of internal radius $ r$ and thickness $\delta r$ is given by:
\begin{equation}
\langle|\tau |^2\rangle^{1/2}=\sqrt{30}\left( \frac{4\pi }5G\right) \left[
\langle a_{2m}(r)^2 \rangle \langle q_{2m}(r)^2 \rangle - \langle 
a_{2m}(r)q_{2m}^{*}(r)\rangle^2\right] ^{1/2}
\label{eq:tau}
\end{equation}
where $q_{lm}$, the multipole moments of the shell and $a_{lm}$, the tidal
moments, are given by:
\begin{equation}
\langle q_{2m}(r)^2 \rangle =\frac{r^4}{\left( 2\pi \right) ^3}M_{sh}^2 
\int k^2dkP\left(k\right) j_2\left( kr\right) ^2
\end{equation}
\begin{equation}
\langle a_{2m}(r)^2 \rangle = \frac{2\rho _b^2r^{-2}}\pi \int dkP
\left( k\right) j_1\left(kr\right) ^2
\end{equation}
\begin{equation}
\langle a_{2m}(r)q_{2m}^{*}(r) \rangle =\frac r{2\pi ^2}\rho _bM_{sh}
\int kdkP\left(k\right) j_1\left( kr\right) j_2(kr)
\end{equation}
where $ M_{sh}$ is the mass of the shell, $j_1(r) 
$ and $j_2(r)$ are the spherical Bessel function of first and second order
while the power spectrum $ P(k)$ is given by:
\begin{eqnarray}
P(k) = Ak^{-1}\left[ \ln \left( 1+4.164k\right) \right] ^2 \nonumber\\
\left(192.9+1340k+1.599\times 10^5k^2+1.78\times 10^5k^3+3.995\times 
10^6k^4\right) ^{-1/2}
\end{eqnarray}
(Ryden \& Gunn 1987). The normalization constant $ A$ can be obtained
imposing that the mass variance at $8h^{-1}Mpc$, $\sigma _{8}$,
is equal to unity.
Filtering the spectrum on cluster scales, $R_{f}=3h^{-1} Mpc$, 
we have obtained the rms torque, $%
\tau (r)$, on a mass shell using Eq. (\ref{eq:tau}) then we 
obtained the total specific
angular momentum, $h(r,\nu)$, acquired during expansion integrating the torque over
time (Ryden 1988 Eq. 35):
\begin{equation}
h(r,\nu)=\frac 13\left( \frac 34\right) ^{2/3}\frac{\tau_o t_0}{M_{sh}}
\overline{\delta}
_o^{-5/2} \int_0^\pi \frac{\left( 1-\cos \theta \right) ^3}{\left(
\vartheta -\sin \vartheta \right) ^{4/3}}\frac{f_2(\vartheta )}{
f_1(\vartheta )-f_2(\vartheta) 
\frac{\delta _o}{\overline{\delta _o}}}d\vartheta \label{eq:ang}
\end{equation}
where $\tau_o$ $\delta_o$ and $\overline{\delta}_o$
are respectively the torque, the mean overdensity and the mean
overdensity within a sphere of radius $r$
at the current epoch $t_0$.
The functions $f_1(\vartheta )$, $f_2(\vartheta )$ are given by Ryden 
(1988 - Eq. 31):
\begin{equation}
f_1(\theta)=16-16 \cos \theta+\sin^2 \theta-9 \theta \sin \theta
\end{equation}
\begin{equation}
f_2(\theta)=12-12 \cos \theta+3 \sin^2 \theta-9 \theta \sin \theta
\end{equation}
where $\theta$ is a a parameter connected to the time, $t$,
through the following
equation:
\begin{equation}
t=\frac{3}{4} t_0 \overline{\delta}_{o}^{-3/2}(\theta -\sin \theta)
\end{equation}
The mean overdensity within a sphere of radius $r$ , $\overline{\delta }%
(r)$, is given by:
\begin{equation}
\overline{\delta} (r,\nu )=\frac 3{r^3}\int_0^r d x x^2 \delta(x)
\label{eq:ma8}
\end{equation}
In fig. 1 we show the variation of 
$h(r,\nu)$ with the distance $r$ for three values of the peak height 
$\nu $. The
rms specific angular momentum, $h(r,\nu)$, increases with distance $r$ while peaks
of greater $\nu $ acquire less angular momentum via tidal torques.
\begin{figure}[ht]
\psfig{file=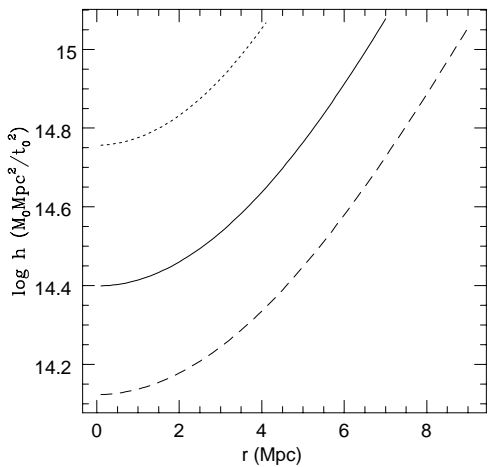,width=11cm}
\caption[]{The specific angular momentum, in units of $M_{\odot}$, Mpc and 
the Hubble time, $t_{o}$, for
three values of the parameter $\nu$ ($\nu=2$ dotted line, $\nu=3$
solid line, $\nu=4$ dashed line)  and for $R_{f}=3h^{-1}Mpc$.}
\end{figure}
This is the angular momentum-density anticorrelation showed by Hoffman
(1986). This effect arises because the angular momentum is proportional to
the gain at turn around time, $t_{m}$, which in turn is proportional
to $\overline{\delta}(r,\nu)^{-\frac 32} \propto \nu^{-3/2}$.

\section{Non-radial motions and the density critical threshold.}

One of the consequences of the angular momentum acquisition by a mass shell
of a proto-cluster is the delay of the collapse of the proto-structure. As
shown by Barrow \& Silk (1981) and Szalay \& Silk (1983) the gravitational
interaction of the irregular mass distribution of proto-cluster with the
neighbouring proto-structures gives rise to non-radial motions, within the
protocluster, which are
expected to slow the rate of growth of the density contrast and to delay or
suppress collapse. According to Davis \& Peebles (1977) the kinetic energy
of the resulting non-radial motions at the epoch of maximum expansion
increases so much to oppose the recollapse of the proto-structure. Numerical
N-body simulations by Villumsen \& Davis (1986) showed a tendency to
reproduce this so called previrialization effect. In a more recent paper by
Peebles (1990) the slowing of the growth of density fluctuations and the
collapse suppression after the epoch of the maximum expansion were re-obtained
using a numerical action method.
In the central regions of a density peak ($r\leq 0.5R_f$) the 
velocity dispersion attain nearly the same value (Antonuccio \& 
Colafrancesco 1997) while at larger radii ($r \geq R_f$) the radial 
component is lower
than the tangential component. This means that motions in the outer regions
are predominantly non-radial and in these regions the fate of the infalling
material could be influenced by the amount of tangential velocity relative
to the radial one. 
This can be shown
writing the equation of motion of a spherically symmetric mass distribution
with density $n(r)$:
\begin{equation}
\frac \partial {\partial t}n \langle v_r \rangle +\frac \partial {\partial r}n 
\langle v_r^2 \rangle + \left(2 \langle v_r^2 \rangle - 
\langle v_\vartheta ^2 \rangle \right) \frac nr+n(r)\frac \partial {\partial
t} \langle v_r \rangle = 0  
\label{eq:peeb}
\end{equation}
where $ \langle v_r \rangle$ and $ \langle v_\vartheta \rangle $ 
are, respectively, the mean radial and
tangential streaming velocity. Eq. (\ref{eq:peeb}) shows that high
tangential velocity dispersion 
$(\langle v_\vartheta ^2 \rangle \geq 2 \langle v_r^2 \rangle)$ 
may alter the infall pattern. The expected delay in the collapse of a 
perturbation may be calculated using a model due to Peebles (Peebles 1993).\\
We consider an ensemble of gravitationally growing mass concentrations,
we suppose that the material in each system collects within the
same potential well
with inward pointing acceleration given by $g(r,t)$. We
indicate with $dP=f(L,r v_r,t)dL dv_r dr$ the probability that a particle
can be found  in the proper radius range $r$, $r+dr$, in the radial
velocity range $v_r={\dot r}$, $v_r+d v_r$ and with angular momentum
$L=r v_\theta$ in the range $dL$.
The radial
acceleration of the particle is: 
\begin{equation}
\frac{dv_r}{dt}=\frac{L^2(r,\nu )}{M^2r^3}-g(r)  
\label{eq:coll}
\end{equation}
where $g(r)$ is the acceleration.
Eq. (\ref{eq:coll}) can be derived from a potential
and then from Liouville's theorem follows that
the distribution function, $f$,
satisfies the collisionless Boltzmann equation:
\begin{equation}
\frac{\partial f}{\partial t} + v_{r}
\frac{\partial f}{\partial r} + \frac{\partial f}{\partial v_{r}}
\cdot \left[ \frac{L_{2}}{r^{3}} - g(r) \right] = 0
\end{equation}
Using Gunn \& Gott's (1972) notation we write the proper radius of a shell in terms of the expansion parameter, $%
a(r_i,t)$, where $r_i$ is the initial radius: 
\begin{equation}
r(r_i,t)=r_{i}a(r_i,t)
\label{eq:ma6}
\end{equation}
and remembering that $M=\frac{4\pi }3\rho (r_i,t)a^3(r_i,t)r_i^3$, 
that $\frac{%
3H_i^2}{8\pi G}=\rho_{ci}$, where $\rho_{ci}$ and $H_i$ are respectively 
the critical mass density and the Hubble constant at the time $t_i$, and assuming that
no shell crossing occurs so that the total mass inside each shell remains
constant, ($\rho (r_i,t)=\frac{\rho _i(r_i,t_i)}{a^3(r_i,t)}$) Eq. (\ref
{eq:coll}) may be written as: 
\begin{equation}
\frac{d^2a}{dt^2}=-\frac{H_i^2(1+\overline{\delta })}{2a^2}+\frac{4G^2L^2}{%
H_i^4(1+\overline{\delta })^2r_i^{10}a^3}  
\label{eq:sec}
\end{equation}
where $\overline{\delta }=\frac{\rho _i-\rho_{ci}}{\rho _{ci}}$, 
or integrating the equation once more: 
\begin{equation}
(\frac{da}{dt})^2=H_i^2\left[ \frac{1+\overline{\delta }}a\right] +\int 
\frac{8G^2L^2}{H_i^4 r_i^{10}\left( 1+\overline{\delta }\right) ^2}\frac
1{a^3}da-2C  
\label{eq:ses}
\end{equation}
where $C$ is the binding energy of the shell. 
Integrating once more we have:
\begin{equation}
t_{ta}=\int_0^{a_{\max }}\frac{da}{\sqrt{H_i^2\left[ \frac{1+\overline{\delta} }a-
\frac{%
1+\overline{\delta }}{a_{\max }}\right] +\int_{a_{\max }}^a\frac{8G^2L^2}{%
H_i^4r_i^{10}(1+\overline{\delta} )^2}}a^3}
\label{eq:sell}
\end{equation}
Using Eqs (\ref{eq:ses}) and (\ref{eq:sell}) it is possible
to find the linear over-density at the turn-around epoch, $t_{ta}$.
In fact solving Eq. (\ref{eq:sell}),
for some epoch of interest, we may obtain the expansion parameter
of the turn-around epoch. This is related to the binding energy of the shell
containing mass $M$ by Eq. (\ref{eq:ses}) with $\frac{da}{dt}=0$.
In turn the binding energy of a growing mode solution is uniquely given
by the linear overdensity, $\delta_{i}$, at time $t_{i}$.
From this overdensity, using linear theory, we may obtain that of
the turn-around epoch.
We find the binding energy of the shell, $C$, using the
relation between $v$ and $\delta_{i}$ for the growing mode
(Peebles 1980) in Eq. (\ref{eq:ses}) and finally the
linear overdensity at the time
of collapse:
\begin{equation}
\delta _c(\nu )=\delta _{co}\left[ 1+\frac{8G^2}{\Omega
_o^3H_0^6r_i^{10}\overline{\delta} (1+\overline{\delta} )^2}\int_0^{a_{\max }}\frac{L^2 \cdot da}{a^3}%
\right]
\label{eq:ma7} 
\end{equation}
where $\delta _{co}=1.68$ is the critical threshold for a spherical model,
while $H_0$ and $\Omega_0$ are respectively
the Hubble constant and the density parameter at the current epoch $t_0$.
We may find the angular momentum, $L$, needed to calculate
$\delta_c$ using Eq.(\ref{eq:ang}).

The mass dependence of the threshold parameter, $ \delta_{c}(\nu)$, can be
found as follows: we calculate the binding radius, $r_{b}$, of the shell 
using Hoffmann \& Shaham's criterion (1985): 
\begin{figure}[ht]
\psfig{file=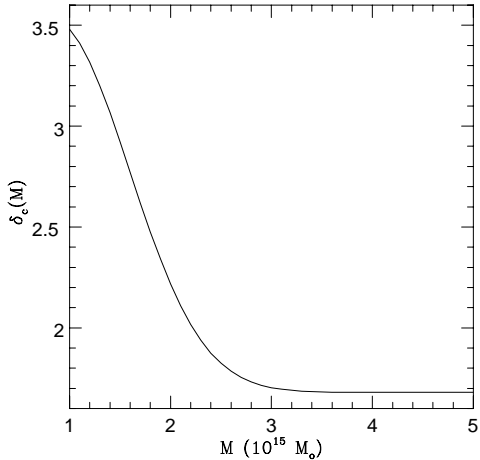,width=11cm}
\caption[]{The threshold $\delta_{c}$ in function of the mass M, for a CDM
spectrum ($\Omega_0=1$, $h=1/2$)
with $R_{f}=3 h^{-1}Mpc$, taking account of non-radial motions.}
\end{figure}
\begin{equation}
T_{c}(r, \nu) \leq t_{0}
\label{eq:ma9}
\end{equation}
where $T_{c}(r,\nu)$ is the calculated time of collapse of a shell and $
t_{o}$ is the Hubble time. We find a relation between $ \nu$ and $M$ through
the equation $ M=4 \pi\rho_b r^{3}_{b}/3$. We so obtain $ \delta_{c}(\nu(M))$. 
In Fig. 2 we show the variation of the threshold parameter, $\delta _c(M)$,
with the mass $M$. Non-radial motions influence the value of $\delta _c$
increasing its value for peaks of low mass while
leaving its value unchanged for
high mass peaks. As a consequence, the
structure formation by low mass peaks is
inhibited. In other words, in agreement with the cooperative
galaxy formation theory (Bower et al. 1993), structures form more
easily in over-populated regions.

\section{Tidal field and the selection function}

According to biased galaxy formation theory the sites of formation of
structures of mass $\sim M$ must be identified with the maxima of the
density peak smoothed over a scale $R_f$ ($M\propto R_f^3$). A necessary
condition for a perturbation to form a structure is that it goes nonlinear
and that the linearly extrapolated density contrast reaches the value 
$\overline{\delta}
(r)\geq \delta _c=1.68$ 
or equivalently that the threshold criterion 
$\nu _t>\delta _c/\sigma_o(R_f)$ is satisfied, being $\sigma_o(R_f)$ 
the variance of the density field smoothed on scale $R_f$.
When these condition are satisfied the matter in a shell around a peak
falls in toward the cluster center and virializes. In this scenario only
rare high $\nu $ peaks form bright objects while low $\nu $ peaks ($\nu
\approx 1$) form under-luminous objects. The kind of objects that form from
nonlinear structures depends on the details of the collapse. Moreover if
structures form only at peaks in the mass distribution they will be more
strongly clustered than the mass. Several feedback mechanisms has been
proposed to explain this segregation effect (Rees 1985, Dekel \& Rees 1987). 
Even if these feedback mechanisms work one cannot expect they have effect
instantaneously, so the threshold for structure formation cannot be sharp 
(BBKS). To
take into account this effect BBKS
introduced a threshold or selection
function, $t(\nu /\nu _t)$. The selection 
function, $t(\nu /\nu _t)$, gives the probability that a
density peak forms an object, while the threshold level, $\nu _t$,
is defined so that the probability that a peak form an object is 1/2 when $
\nu =\nu _t$. 
The selection function introduced by BBKS (Eq. 4.13), is an empirical one
and depends on two parameters: the threshold $\nu _{t}$ and the shape
parameter $ q$:
\begin{equation}
t(\nu /\nu _t)=\frac{(\nu /\nu _t)^q}{1+(\nu /\nu _t)^q}
\end{equation}
If $q\rightarrow \infty $ this selection function is a Heaviside function $%
\vartheta (\nu -\nu _t)$ so that peaks with $\nu >\nu _t$ have a
probability equal to 100\% to form objects while peaks 
with $ \nu \leq \nu_{t}$ do not form objects. 
If $q$ has a finite value 
sub-$\nu _t$ peaks are selected with non-zero probability. Using the given
selection function the cumulative number density of peaks higher than $\nu $
is given, according to BBKS, by:
\begin{equation}
n_{pk}=\int_\nu ^\infty t(\nu /\nu _t)N_{pk}(\nu )d\nu 
\end{equation}
where $N_{pk}(\nu )$ is the comoving peak density 
(see BBKS Eq.~4.3).
A form of the selection function, physically motivated, can be obtained
following the argument given in CAD. 
In this last paper the selection function is defined as:
\begin{equation}
t(\nu )=\int_{\delta _c}^\infty p\left[ \overline{\delta} ,
\langle \overline{\delta} \rangle (r_{Mt},\nu
),\sigma _{\overline{\delta}} (r_{Mt},\nu )\right] d\delta \label{eq:sel}
\end{equation}
where the function 
\begin{equation}
p\left[ \overline{\delta} ,\langle \overline{\delta} \rangle (r)\right] = 
\frac 1{\sqrt{2\pi }\sigma_{\overline{\delta}} }\exp \left( -\frac{|\overline{\delta} -
\langle \overline{\delta} \rangle (r)|^2}{2\sigma
_{\overline{\delta}} ^2}\right) \label{eq:gau}
\end{equation}
gives the probability that the peak overdensity is
different from the average, in a Gaussian density field. The selection
function depends on $\nu $ through the dependence of $\overline{\delta} (r)$ from $\nu $. 
As displayed the integrand is evaluated at a radius $r_{Mt}$ which is the
typical radius of the object that we are selecting. Moreover the selection
function $t(\nu )$ depends on the critical overdensity threshold for the
collapse, $\delta _{c}$, which is not constant as in a spherical model
(due to the presence, in our analysis, of non-radial motions that delay the
collapse of the proto-cluster) but it depends on $\nu $. 
Known $\delta_{c}(\nu)$ and chosen a spectrum, the selection function is
immediately obtainable through Eq. (\ref{eq:sel}) and Eq. (\ref{eq:gau}).
The result of the calculation, plotted in Fig. 3, for two values
of the filtering radius, ($R_{f}=2$, $3$ $h^{-1}Mpc$),
shows that the selection
function, as expected, differs from an Heaviside function (sharp threshold).
\begin{figure}[ht]
\psfig{file=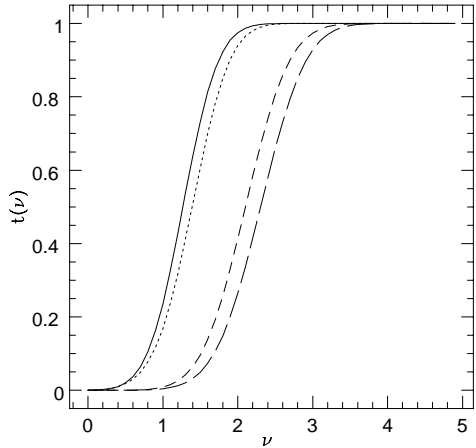,width=11cm}
\caption[]{The selection function, $t(\nu)$, for $R_{f}= 3h^{-1}Mpc$
($\delta_{c}=1.68$, solid line; $\delta_{c}$ function of $\nu$, dotted line) 
and for $4h^{-1}Mpc$ ($\delta_{c}=1.68$, short dashed line;
$\delta_{c}$ function of $\nu$, long dashed line).}
\end{figure}
The value of $\nu$ at which the selection function $ t(\nu)$ reaches the 
value 1 ($t(\nu) \simeq 1$) increases for growing values of the filtering
radius, $R_{f}$. This is due to the smoothing effect of the filtering process.
The effect of non-radial motions is, firstly, that of shifting
$t(\nu)$ towards higher values of $\nu$, and, secondly, that of
making it steeper. 
The selection function is also different from that
used by BBKS (Tab. 3a). 
Finally it is interesting to note that the selection function defined by
Eq. (\ref{eq:sel}) and Eq. (\ref{eq:gau}) is totally general, it does not
depend on the presence or absence of non-radial motions. The latter
influence the selection function form through the changement
of $\delta_{c}$ induced by non-radial motions itself.

\section{The bias coefficient}

A model of the Universe in which light traces the mass distribution
accurately (unbiased model) is subject to several problems. As pointed
out by Davis et al. (1985) an unbiased CDM produces a galaxy correlation
function which is steeper than observed and a pairwise velocity dispersion
larger than that deduced from redshift surveys. A remedy to this
problem can be found if the assumption that light trace mass is relaxed
introducing the biasing concept, i.e. that galaxies are more clustered than
the distribution of matter in agreement to the concept of biasing inspired
by Kaiser's (1984) discussion of the observation that clusters of galaxies
cluster more strongly than do galaxies, in the sense that the dimensionless
two-point correlation function, $\xi _{cc}(r)$, is much larger than the
galaxy two-point function, $ \xi _{gg}(r)$.
The galaxy two-point correlation function $ \xi _{gg}(r)$
is a power-law :
\begin{equation}
\xi_{g}(r) = (\frac{r}{r_{0,g}})^{\gamma}
\label{eq:unoventotto}
\end{equation}
with a correlation length $ r_{0,g} \simeq 5 h^{-1}$ Mpc and
a slope $ \gamma \simeq 1.8$ for
$ r \le 10 h^{-1}$ Mpc (Davis \& Peebles 1983; Davis et al. 1985; 
Shanks et al. 1989), (some authors disagree with this values; 
for example Strauss et al. 1992 and Fisher et al. 1993  
find $ r_{0,g} \simeq 3.79 h^{-1}$ Mpc and $ \gamma \simeq 1.57$). 
As regards the clusters of galaxies the form of the 
two-point correlation function, $\xi _{cc}(r)$, is equal to that given
by Eq. (\ref{eq:unoventotto}). Only the correlation length
is different. In the case
of clusters of galaxies the value of $ r_{0,c}$ is uncertain (see 
Bahcall \& Soneira 1983; Postman et al. 1986; Sutherland 1988; 
Bahcall 1988; Dekel et al. 1989; Olivier et al. 1990  and Sutherland \& 
Efstathiou 1991) however it lays in the range 
$  r_{0,c} \simeq 12 \div 25 h^{-1}$ Mpc in any case larger than $ r_{0,g}$.
One way of defining the bias coefficient of a class of objects
is that given by (BBKS):
\begin{equation}
b(R_f)= \frac { \langle \tilde{\nu} \rangle } {\sigma_0}+1
\end{equation}
where $ \langle \tilde{\nu} \rangle$ is:
\begin{equation}
\langle \tilde{\nu} \rangle = \int_0^\infty \left[ \nu -\frac{\gamma \theta }{1-\gamma ^2}\right] 
t(\frac{\nu}{\nu_{t}}) N_{pk}(\nu ) d\nu 
\label{eq:nu}
\end{equation}
from Eq. (\ref{eq:nu})
it is clear that the bias parameter can be calculated once a
spectrum, $P (k)$, is fixed. The bias parameter depends on the shape 
and normalization of the power spectrum. A larger value
is obtained
for spectra with more power on large scale (Kauffmann et al. 1996). 
In this calculation we continue to use the standard CDM
spectrum ($\Omega_0 = 1$, $ h = 0.5$) normalized imposing that the rms density
fluctuations in a sphere of radius $ 8 h^{-1}Mpc$ is the same as that observed
in galaxy counts, i.e. $\sigma _8=\sigma (8h^{-1}Mpc)=1$. The calculations
have been performed for three different values of the
filtering radius ($R_f=2,$ $ 3$, $ 4$
$h^{-1}Mpc$). The values of $ b$, that we have obteined, are respectively, 
in order growing of $ R_f$, 1.6, 1.93 and 2.25.

As shown,
the value of the bias parameter tends to increase with $R_f$ due the filter
effect of $ t(\nu)$. As shown $t(\nu)$ acts as a filter, increasing the
filtering radius, $R_{f}$, the value of $\nu$ at which
$t(\nu) \simeq 1$ increases . In other words when $ R_{f}$ increases $t(\nu)$ selects
density peaks of larger height. The reason of this behavior must be
searched in the smoothing effect that the increasing of the filtering radius
produces on density peaks. When $ R_{f}$ is increased the density field
is smoothed and $t(\nu)$ has to shift towards higher value of $ \nu$
in order to select a class of object of fixed mass,  $M$. 




\section{The X-ray temperature function}

The PS theory provides an analytical description of the
evolution of structures in a hierarchical Universe. In this model the linear
density field, $\rho ({\bf x},t)$, is an isotropic random Gaussian field, the
non-linear clumps are identified as over-densities (having a density contrast 
$\delta _c\sim 1.68$ - Gunn \& Gott 1972) in the linear density field, while
a mass element is incorporated into a non-linear object of mass $M$ when the
density field smoothed with a top-hat filter of radius $R_f$, exceeds a
threshold $\delta _c$ ($ M\propto R_f^3$).
The probability distribution for
fluctuations is given by:
\begin{equation}
p[\delta (M)]=\frac 1{\sqrt{2\pi }\sigma (M)}\exp [-\delta (M)^2/2\sigma
(M)^2] \label{eq:pip}
\end{equation}
The probability that an object of mass $M$ has formed is obtained
integrating Eq. (\ref{eq:pip}) from the threshold value $\delta _c$ to
infinity and the
comoving number density of non-linear objects of mass $M$ to $M+dM$ is given
simply by differentiating the integral with respect to mass and is given by:
\begin{equation}
N(M,t)dM=-\rho_b \sqrt{\frac 2\pi }\nu \exp \left( -\nu ^2/2\right) \frac
1\sigma \left( \frac{d\sigma }{dM}\right) \frac{dM}M  
\label{eq:press}
\end{equation}
where $\rho_b$ is the mean mass density, $\sigma (M)$ is the rms linear mass
overdensity
evaluated at the epoch when the mass function is desidered and $\nu =%
\frac{\delta _c}{\sigma (M)}$. The redshift dependence of
Eq. ~(\ref{eq:press}) can be
obtained remembering that
\begin{equation}
\nu =\frac{\delta _c(z)D(0)}{\sigma _o(M)D(z)}
\end{equation}
being $ D(z)$ the growth factor of the density perturbation and
$\sigma_o(M)$ the current value of $\sigma (M)$.
In Eq. ~(\ref{eq:press})
PS introduced arbitrarily a factor of two because 
$ \int_0^\infty dF(M)=1/2$, so that only half of the mass in the Universe
is accounted for. Bond et al. (1991) showed that the "fudge factor" 2 is
naturally obtained using the excursion set formalism in the sharp
$k$-space while for general filters (e.g., Gaussian or "top hat")
it is not possible to obtain an analogous analytical result. As stressed by
Yano et al. 1996, the factor of 2 obtained in the sharp $k$-space
is correct only if the spatial correlation of the density fluctuations
is neglected.
In spite of the quoted problem, several authors (Efstathiou et al.
1988; Brainerd \& Villumsen 1992; Lacey \& Cole 1994) showed that PS analytic
theory correctly agrees with N-body simulations.
In particular Efstathiou et al. (1988),
showed
that PS theory correctly agrees with                 
the evolution of the distribution of mass amongst groups and clusters of
galaxies (multiplicity function). Brainerd \& Villumsen (1992) studied the
CDM halo mass function using a hierarchical particle mesh code. From this
last work results that PS formula fits the results of the simulation up
to a mass of ~10 times the characteristic 1$\sigma$ fluctuation mass,
$M_{\ast}$, being $M_{\ast} \simeq 10^{15} b^{-6/(n_{l}+3)} M_{\odot}$,
where $b$ is the
bias parameter 
and $n_{l}$ is the local slope of the power spectrum.
PS theory has proven particularly useful in
analyzing the number counts and redshift distributions for QSOs (Efstathiou
\& Rees 1988), Lyman $\alpha $ clouds (Bond et al. 1988) and
X-ray clusters (Cavaliere \& Colafrancesco 1988). \\
Some difficulties arise when PS theory is compared with observed
distributions. To estimate the multiplicity function of real systems it
is in fact required a knowledge of the temperature-mass (T-M) relation in
order to trasform the mass distribution into the temperature distribution.
Theoretical uncertainty arises in this transformation because the exact
relation between the mass
appearing in the PS expression and the temperature of the intracluster
gas is unknown.
Under the standard assumption of the 
Intra-Cluster (IC) gas in hydrostatic equilibrium with the potential well 
of a spherically simmetric, virialized cluster, the IC gas temperature-mass
relation is easily obtained by applying the virial
theorem
and for a flat matter-dominated Universe
we have that (Kaiser 1986, Evrard 1990):
\begin{equation}
T=(6.4h^{2/3}keV)\left( \frac M{10^{15}M_{\odot }}\right) ^{2/3}(1+z) 
\label{eq:ma2}
\end{equation}
The assumptions of perfect hydrostatic equilibrium and virialization
are in reality non completely satisfied by clusters. Clusters profile may
have departure from isothermality, with sligth temperature gradients
throughout the cluster. The X-ray weighted temperature can be slightly
different from the mean mass weighted virial temperature. In any
case
the scatter in the T-M relation given by Eq. (\ref{eq:ma2})
is of the order of 
$\simeq 10 \%$ (Evrard 1991).
As shown by Bartlett \& Silk (1993) the X-ray distribution
function obtained using a standard CDM spectrum over-produces the clusters
abundances data obtained from Henry \& Arnaud (1991) and Edge et al.
(1990). The
discrepancy can be reduced taking into account the non-radial motions that
originate when a cluster reaches the non-linear regime.
In fact, the PS temperature distribution requires specification
of $\delta _c$ and the temperature-mass relation T-M. The presence of non-radial
motions changes both $\delta _c$ and the T-M relation.  
To get the temperature distribution it is necessary to know 
the temperature-mass
relation. This can be obtained using the virial theorem, energy
conservation and using Eq. (\ref{eq:ses})
(Bartlett \& Silk 1993).
From the virial theorem we may write:
\begin{equation}
\langle K \rangle = \frac{GM}{2 r_{eff}}+\int_{0}^{r_{eff}} \frac{L^2}{2 M^2 r^3} d r
\label{eq:vir}
\end{equation}
while from the energy conservation:
\begin{equation}
- \langle K \rangle +\frac{GM}{r_{eff}}+\int_{0}^{r_{eff}} \frac{L^2}{M^2 r^3} d r=
\frac{GM}{r_{ta}}+\int_{0}^{r_{ta}} \frac{L^2}{M^2 r^3} d r
\label{eq:ener}
\end{equation}
Eq. (\ref{eq:vir}) and Eq.(\ref{eq:ener}) can be solved for
$r_{eff}$ and $<K>$.
We finally have that:
\begin{equation}
T=(6.4 keV)\left( \frac{M \cdot h}{10^{15} M_{\odot }}\right)^{2/3}
\left[ 1+
\frac{\eta \psi \int_0^r \frac{L^2 dr}{M^2 r^3}}{(G^{2}
\frac{H_0^2\Omega_0}{2} M^{2})^{1/3}}\right] 
\label {eq:temp} 
\end{equation}
where
$\eta$ is a parameter given by $ \eta=r_{ta}/x_{1}$, being $r_{ta}$
the radius of the turn-around epoch, while $x_{1}$ is defined
by the relation
$M=4 \pi \rho_b x^{3}_{1}/3$ and $ \psi=r_{eff}/r_{ta}$ where $r_{eff}$
is the time-averaged radius of a mass shell.
Eq. (\ref{eq:temp}) was normalised to agree with Evrard's (1990) simulations
for $L=0$. \\
The new T-M relation, Eq. (\ref{eq:temp}), differs from 
Eq. (\ref{eq:ma2}) for the presence of the term:
\begin{equation} 
\frac{\eta \psi \int_0^r \frac{L^2dr}{M^2r^3}}{(G^{2}%
\frac{H_0^2\Omega_0}{2} M^{2})^{1/3}} \nonumber
\label{eq:diff}
\end{equation}
This last term changes the dependence of the temperature from
the mass, $M$, in the T-M relation. Moreover the new T-M relation
depends on the angular momentum, $L$,  
originating from the gravitational interaction of the quadrupole moment
of the protocluster with the tidal field of the matter of the
neighboring protostructures.
In Fig. 4 the X-ray temperature distribution, derived using
a CDM model with $\Omega_0=1$, $h=1/2$ and taking into account of
non-radial motions, is compared with Henry \& Arnaud (1991) and Edge et al.
(1990) data and with a pure CDM model with $\Omega_0=1$, $h=1/2$.
\begin{figure}[ht]
\psfig{file=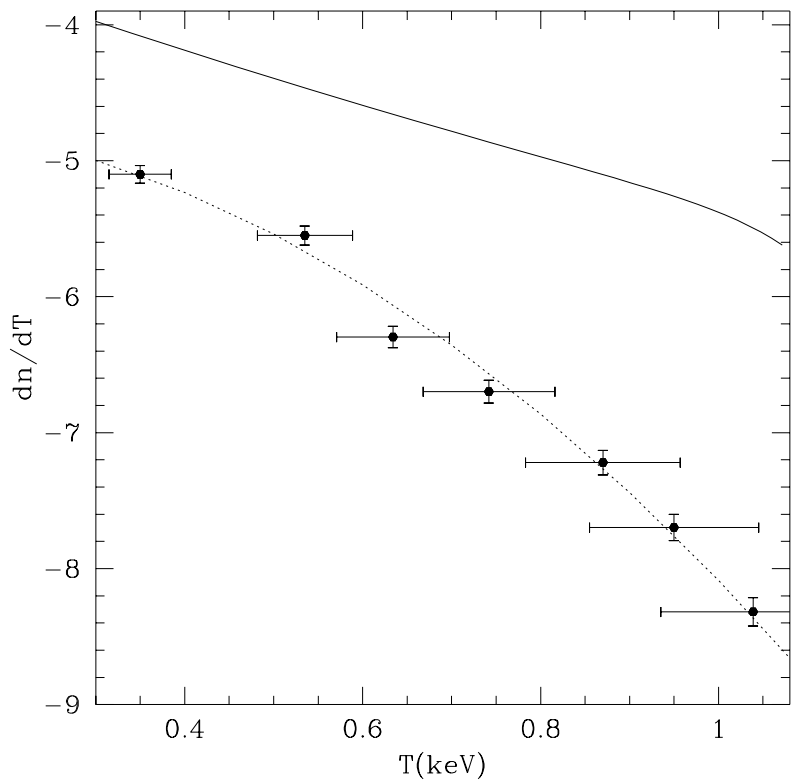,width=11cm}
\caption[]{X-ray temperature distribution function. The solid line gives the
temperature function for a pure CDM model ($\Omega_0=1$, $h=1/2$),
with $ R_{f} = 3 h^{-1} Mpc$. The dotted
line is the same distribution but now taking account of non-radial motions. 
The data are obtained by Edge et al. 1990, and Henry \& Arnaud 1991}
\end{figure}
As shown the CDM model that does not take account of the non-radial
motions over-produces the clusters abundance.  
The introduction of non-radial motions
gives a more careful description of the experimental data. 
As we have seen
the X-ray temperature distribution function obtained taking account
of non-radial motions is different from that of a pure CDM model for two
reasons: \\
1) the variation of the threshold, $\delta_c$, with mass, $M$.
This is due to the changement of the energetics of the
collapse model
produced by the introduction of another potential energy term
($\frac{L(r,\nu)^2}{M^2 r^3}$) in
$Eq. (\ref{eq:coll})$;  \\
2) the modification of the T-M relation produced
by the alteration of the partition of energy in virial equilibrium. \\
For values of mass $M=0.5 M_{\odot}$ the difference between
the two theoretical lines in Fig. 4 is due to the first factor
for a $\simeq 59 \%$ and this value increases with increasing mass. 
The uncertainty in our model 
fundamentally comes from the uncertainty of the T-M relation whose
value has been previously quoted.\\
One of the objection to the result of this paper may that
the effect described has not been seen in some hydrodynamic simulations
(Evrard \& Crone 1992). The answer to this objection is 
that our model is fundamentally based on the previrialization
conjecture (Davis \& Peebles 1977; Peebles 1990), (supposing that initial
asphericities and tidal interactions between neighboring density fluctuations
induce significant non-radial motions, which oppose the collapse) and while
some N-body simulations (Villumsen \& Davis 1986; Peebles 1990)
appear to reproduce this effect, other simulations
(for example Evrard \& Crone 1992) do not. An answer to this controversy
was given by (Lokas et al. 1996). The problem is connected to spectral
index $n$ used in the simulations. The "previrialization"
is seen only for $n>-1$. While Peebles (1990) used simulations with $n=0$,
Evrard \& Crone (1992) assumed $n=-1$.
Excluding this particular case generally the ensemble properties of
clusters like their optical and X-ray luminosity functions, or their
velocity and temperature distribution functions, are difficult to address
directly in numerical simulations because the size of the box must be very
large in order to contain a sufficient number of clusters; an then analytical
approach remains an effective alternative.\\

\section{Non-radial motions and the clusters correlation function}

Supposing that the Universe at some early epoch can be described using 
the same assumptions of the previous section 
the probability $p(M_1,M_2,r)$ per unit volume
per unit masses of finding two collapsed objects of mass $M_1$ and $M_2$
separated by a distance $r$ is obtained by integrating both variables of the
bivariant Gaussian distribution in $\delta (M_1)$ and $\delta (M_2)$, with
correlation $\xi _\rho (r)$ from
$\delta _c$ to infinity and then taking the
partial derivatives with respect to both masses. The correlation function
for collapsed objects is simply obtained from:
\begin{equation}
\xi _{MM}=p(M_1,M_2,r)/p(M_1)p(M_2)-1 
\end{equation}
\noindent which for equal masses and weak correlations $\xi _\rho \ll 1$ is
(Kashlinsky 1987):
\begin{equation}
\xi _{MM}=[\delta _c^2/\sigma _\rho (M)^2]\xi _\rho (r)  \label{eq:coo}
\end{equation}
where $ \sigma _\rho (M)$ is the variance of the mass fluctuation and
$\xi_{\rho}$ is the correlation function of the matter density
distribution when the density fluctuations had small amplitude.
Eq. (\ref{eq:coo}) shows that the correlation of collapsed objects may be
enhanced relative to that of the underlying mass fluctuations. This condition
is usually described by the bias parameter $b$ which is sometimes defined
as $[\xi_{MM}(r)/\xi_{\rho}(r)]^{1/2}=\delta_{c}/\sigma_{\rho}(M)$.\\
Studies of clustering on scales $\geq 10h^{-1}Mpc$ have shown that
the correlation function given in Eq. (\ref{eq:coo}) is different from
that obtained from
observations. The most compelling data are angular correlation functions for
the APM survey. These decline much less rapidly on large scales than the CDM
prediction (Maddox et al. 1990). As discussed in the introduction there are
two ways to reduce the quoted discrepance: either with a modification of 
CDM theory involving the physics of early Universe, or with a modification 
of CDM theory involving the physics of galaxy formation.
This discrepancy can be reduced, similarly to the problem 
of over-abundance of clusters, taking into
account the non-radial motions that originate when a cluster reaches the
non-linear regime. In fact, the calculation  of the correlation function 
requires the specification of $\delta _c$ which is changed by non-radial 
motions and whose new value is given by Eq. (\ref{eq:ma7}).
In Fig. 5 the correlation function derived taking into account
non-radial motions is compared both with the two-point correlation
function obtained by Sutherland \& Efstathiou (1991) from the analysis
of Huchra's et al. (1990) deep redshift survey as discussed in Geller 
\& Huchra (1988) and with the data points for the APM clusters 
computed by Efstathiou et al. (1992).
\begin{figure}
\psfig{file=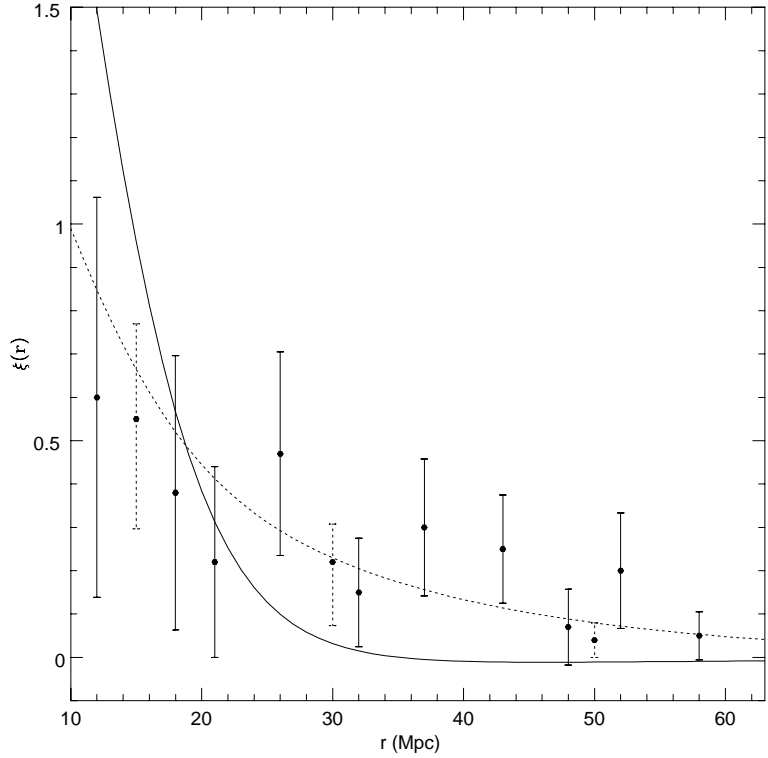,width=11cm}
\caption[]{Clusters of galaxies correlation function.
The solid line gives the
correlation function for a pure CDM model, with $R_{f}=3 h^{-1} Mpc$. The dashed
line is the same distribution but now taking account of non-radial motions. 
The observational data refer to the two-point correlation
function obtained by Sutherland \& Efstathiou (1991) ({\it filled  exagons})
from the analysis of Huchra's et al. (1990) deep redshift survey and 
with the data points for the APM clusters computed by Efstathiou et 
al. (1992) ({\it dashed errorbars}).}
\end{figure}
As shown the discrepancy between pure CDM previsions and
experimental data is remarkable. The CDM model seems to have
trouble in re-producing the behaviour of the data. In fact, the predicted
two-point cluster function is too steep and rapidly goes nearly to zero for
$r \simeq 30 h^{-1} Mpc$, while the data show no significant anticorrelation
up to   $r \simeq 60 h^{-1} Mpc$ (see Borgani 1990).
The introduction of non-radial motions
gives a more accurate description of the experimental data.
The result obtained is in agreement with that of Borgani (1990) who studied
the effect of particular thresholds (erfc-threshold and Gaussian-threshold)
on the correlation
properties of clusters of galaxies.
The fundamental difference between our and Borgani's approach is that our
threshold function is physically motivated: it is simply obtained from the
assumptions of a Gaussian density field and taking account of non-radial
motions. Borgani's threshold functions (erfc and Gaussian threshold) are
ad-hoc introduced in order to reduce the discrepancy between the observed and
the CDM predicted two points correlation functions of clusters of galaxies.
The connection with the quoted non-sphericity effects, even if
logical and in agreement with our results, is only a posteriori tentative
to justify the choice made.

\section{Conclusions}


In these last years many authors
have shown the existence of a
strong discrepancy between the observed properties of clusters 
of galaxies and that predicted by the CDM model. 
To reduce this discrepancy several alternative models have been introduced 
but no model has considered the role of the non-radial motions. Here 
we have shown how non-radial motions may reduce some of these
discrepancies: namely \\
a) the discrepancy between the observed $ \xi(r)$ of cluster of 
galaxies and that predicted by CDM model;\\ 
b) the discrepancy between the measured X-ray temperature distribution 
function and that predicted by the CDM model. \\
To this aim we calculated
the variation of the treshold parameter,
$\delta_{c}$, as a function of the mass $M$, produced by the presence
of non-radial motions in the outskirts of clusters of galaxies.
We used $\delta_{c}(M)$ to calculate the two-point correlation of
clusters of galaxies and the X-ray temperature distribution function. 
We compared the prediction for the two point correlation function with
that obtained
by Sutherland \& Efstathiou (1991) from the analysis
of Huchra's et al. (1990) deep redshift survey as  
discussed in Geller \& Huchra (1988) and with the
data points for the APM clusters computed by Efstathiou et al. (1992). 
The prediction for the X-ray temperature function was compared 
with Henry \& 
Arnaud (1991) and Edge et al. (1990)  
X-ray temperature distributions for local clusters.
Our results (see Fig. 4 and Fig. 5) show how the non-radial motions change 
both the correlation length of the correlation function, making it less steep 
than that obtained from a pure CDM model where the non-radial motions are 
not considered, and the X-ray temperature function. In both cases our model
gives a good agreement with the data.\\
Finally we calculated the bias coefficient using a selection function
that takes into account the effects of non-radial motions, and we show that 
the {\it bias} so obtained can account for a substantial part of the total 
bias required by observations on cluster scales.\\
~\\
~\\

\end{document}